\documentclass[conference,table]{IEEEtran}
\usepackage[utf8]{inputenc}
\usepackage{textgreek}
\IEEEoverridecommandlockouts

\usepackage{bbding}
\usepackage{cite}
\usepackage{amsmath,amssymb,amsfonts}
\usepackage{bm}
\usepackage{textcomp}
\usepackage{caption}
\usepackage{graphicx}
\usepackage{pgfplots}
\pgfplotsset{compat=1.18}
\usepackage{subcaption}
\usepackage{multirow}
\usepackage{adjustbox}
\usepackage{algorithmic}
\usepackage{algorithm}
\usepackage{breqn}
\usepackage{booktabs}
\usepackage{siunitx}
\usepackage{tabularx}
\usepackage{comment}
\usepackage{authblk}
\usepackage{adjustbox}
\usepackage{tikz}
\usetikzlibrary{arrows.meta, positioning, shadows}
\usepackage[skip=3pt]{caption} 
\usepackage{fancyhdr}
\fancypagestyle{firstpage}
{
    \fancyhead[L]{\footnotesize © 2026 IEEE.  Personal use of this material is permitted.  Permission from IEEE must be obtained for all other uses, in any current or future media, including reprinting/republishing this material for advertising or promotional purposes, creating new collective works, for resale or redistribution to servers or lists, or reuse of any copyrighted component of this work in other works. This paper is accepted at the 12th International Conference on Computing and Artificial Intelligence (ICCAI) 2026.}
    \fancyhead[R]{}
}

\begin{document}

\title{DART: Input-\underline{D}ifficulty-\underline{A}wa\underline{R}e Adaptive \underline{T}hreshold for Early-Exit DNNs}

\author[1]{Parth Patne}
\author[1,2]{Mahdi Taheri}
\author[1]{Christian Herglotz}
\author[2]{Maksim Jenihhin}
\author[3]{Milos Krstic}
\author[1]{Michael Hübner}

\affil[1]{Brandenburg Technical University, Cottbus, Germany}
\affil[2]{Tallinn University of Technology, Tallinn, Estonia}
\affil[3]{Leibniz Institute for High Performance Microelectronics, Frankfurt Oder, Germany}

\maketitle
\thispagestyle{firstpage}

\begin{abstract}
Early-exit deep neural networks enable adaptive inference by terminating computation when sufficient confidence is achieved, reducing cost for edge AI accelerators in resource-constrained settings. Existing methods, however, rely on suboptimal exit policies, ignore input difficulty, and optimize thresholds independently. This paper introduces DART (Input-Difficulty-Aware Adaptive Threshold), a framework that overcomes these limitations. DART introduces three key innovations: (1) a lightweight difficulty estimation module that quantifies input complexity with minimal computational overhead, (2) a joint exit policy optimization algorithm based on dynamic programming, and (3) an adaptive coefficient management system. Experiments on diverse DNN benchmarks (AlexNet, ResNet-18, VGG-16) demonstrate that DART achieves up to \textbf{3.3$\times$} speedup, \textbf{5.1$\times$} lower energy, and up to \textbf{42\%} lower average power compared to static networks, while preserving competitive accuracy. Extending DART to Vision Transformers (LeViT) yields power (5.0$\times$) and execution-time (3.6$\times$) gains but also accuracy loss (up to 17 percent), underscoring the need for transformer-specific early-exit mechanisms.
We further introduce the Difficulty-Aware Efficiency Score (DAES), a novel multi-objective metric, under which DART achieves up to a 14.8× improvement over baselines, highlighting superior accuracy, efficiency, and robustness trade-offs.
\end{abstract}

\begin{IEEEkeywords}
Deep Neural Networks, Dynamic Neural Networks, Edge AI, Hardware Accelerators, Efficient Computing
\end{IEEEkeywords}

\section{Introduction}
Dynamic Deep Neural Networks (D2NNs) address the inefficiency of traditional static inference, where every input traverses the full network despite high energy consumption and latency, by adapting computation to input characteristics, making them more suitable for resource-constrained edge accelerators~\cite{han2022dynamic}. Early exit networks, introduced first by BranchyNet~\cite{teerapittayanon2016branchynet}, integrate auxiliary classifiers at intermediate layers, allowing confident predictions to exit early. This approach demonstrated that many inputs require only shallow processing, enabling significant savings. However, reliance on fixed confidence thresholds and independent exit optimization limits adaptability during real-time execution.

Subsequent work has improved exit policies. Zhou et al.~\cite{zhou2020bert} and Taheri et al.~\cite{taheri2025rlagent} employed reinforcement learning to optimize exits, while Xin et al.~\cite{xin2020deebert} introduced DeeBERT for transformer acceleration. RACENet~\cite{racenet} incorporates class-awareness through adaptive normalization but relies on computationally expensive MLPs at every layer, negating efficiency gains on constrained edge devices. 

Despite all efforts in the literature, three critical limitations persist. First, state-of-the-art approaches~\cite{wang2024early,wu2023early} optimize exit thresholds independently, ignoring their interdependencies and yielding suboptimal routing policies. Second, the estimation of input complexity is computationally expensive~\cite{peli1990,forsythe2011predicting} or insufficiently representative of neural processing needs~\cite{lee2020neural}, limiting the real-time deployment. Third, current systems adopt static exit policies learned during training, leaving them vulnerable to distribution shifts and operational variability, with little work exploring online adaptation~\cite{finn2017model,ash2019warm} in different contexts that need to be studied in the D2NN domain.

To overcome these limitations, this paper introduces \textbf{DART} (Input-Difficulty-Aware Adaptive Threshold), a unified framework featuring:
(1) \textbf{Difficulty-aware input processing} for lightweight real-time complexity estimation;
(2) \textbf{Joint exit threshold optimization} using dynamic programming for globally coherent routing; and
(3) \textbf{Adaptive coefficient management} with online learning--based updates for continuous refinement during deployment.

The main contributions of this work are:
\begin{itemize}
\item \textbf{A unified framework (DART)} that integrates difficulty-aware input processing, joint threshold optimization, and adaptive management to overcome fundamental early-exit limitations;
\item \textbf{A practical and extensible deployment methodology} that enables real-time execution on edge accelerators while ensuring portability across diverse neural network architectures;
\item \textbf{Open-source release and comprehensive evaluation} of the fully automated framework on state-of-the-art CNN and vision transformer benchmarks.
\end{itemize}
The remainder of this paper is structured as follows: Section II presents the proposed methodology; Section III describes the experimental setup and results; Section IV concludes the paper.

\begin{figure}[!t]
\centering
\includegraphics[width=0.9\linewidth]{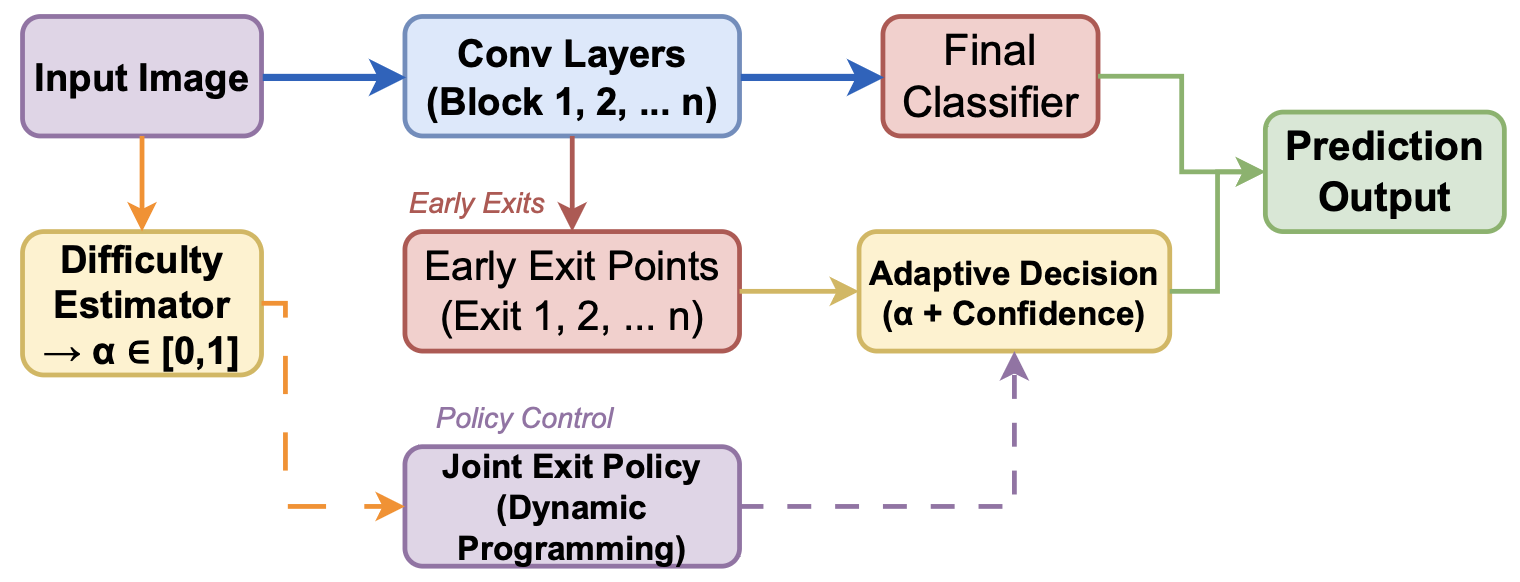}
\caption{System-level architecture of the proposed Input-Difficulty-Aware early-exit DNN with DART integration.}
\label{fig:architecture_intro}
\end{figure}

\section{Methodology}

Figure~\ref{fig:architecture_intro} illustrates the system-level integration of DART into an early-exit DNN, showing how difficulty estimation, adaptive thresholding, and coefficient management are realized within the model pipeline.
DART implements a systematic approach to dynamic neural network optimization through three interconnected components that work synergistically to achieve optimal efficiency--accuracy trade-offs.
Within a DART-enabled model, inputs are pre-processed by the lightweight difficulty-estimation module, exit thresholds are dynamically adjusted using the learned policies, and adaptation is performed online during inference. The framework itself provides the methodology and optimization algorithms that enable these runtime modules.

\subsection{Difficulty-Aware Input Processing}
The foundation of DART lies in its ability to quantify input complexity through a lightweight preprocessing module. This module combines three complementary metrics to provide a comprehensive difficulty assessment.

\subsubsection{Multi-Modal Difficulty Metrics}
\textbf{Edge Density Computation:} Edge density quantifies the structural complexity of input images through gradient analysis. The computation employs Sobel operators in both horizontal and vertical directions:
\begin{align}
G_x(x,y) &= I(x,y) * S_x, \label{eq:sobel_x}\\
G_y(x,y) &= I(x,y) * S_y, \label{eq:sobel_y}\\
G(x,y) &= \sqrt{G_x^2(x,y) + G_y^2(x,y)}, \label{eq:gradient_magnitude}
\end{align}
where \(I(x,y)\) represents the grayscale-converted input image of height $H$ and width $W$, \(S_x\) and \(S_y\) are the Sobel kernels~\cite{Sobel1968}, and \(*\) denotes convolution. The edge density score is computed as:
\begin{equation}
\alpha_{\text{edge}} = \frac{1}{HW}
\sum_{x=1}^{H} \sum_{y=1}^{W} \mathbf{1}\bigl[G(x,y) > \tau_{\text{edge}}\bigr],
\label{eq:edge_density}
\end{equation}
where \(\mathbf{1}[\cdot]\) is the indicator function and \(\tau_{\text{edge}}\) is an adaptive threshold tuned per scene complexity~\cite{peli1990}.

\textbf{Pixel Variance Analysis:} Spatial variance captures texture complexity and local variations within the input. For an input tensor \(X \in \mathbb{R}^{B \times C \times H \times W}\) (where $B$ is batch size, $C$ is channels, $H$ is height, and $W$ is width), the variance is computed across spatial dimensions:
\begin{align}
\mu_{b,c} &= \frac{1}{HW}
\sum_{h=1}^{H} \sum_{w=1}^{W} X_{b,c,h,w}, \label{eq:mean}\\
\alpha_{\text{variance}} &= \frac{1}{CHW}
\sum_{c=1}^{C} \sum_{h=1}^{H} \sum_{w=1}^{W}
\bigl(X_{b,c,h,w} - \mu_{b,c}\bigr)^2, \label{eq:variance}
\end{align}
following established texture complexity analyses.

\textbf{Gradient Complexity Assessment:} Second-order spatial variations are captured using Laplacian operators to detect fine-grained patterns:
\begin{equation}
\alpha_{\text{gradient}} = \frac{1}{HW}
\sum_{x=1}^{H} \sum_{y=1}^{W} \lvert L(x,y)\rvert,
\quad
L(x,y) = I(x,y) * K_{\text{laplacian}},
\label{eq:laplacian}
\end{equation}
where \(K_{\text{laplacian}}\) is the Laplacian kernel, reflecting classic second-order edge detection.

\subsubsection{Difficulty Score Fusion}
The final difficulty score combines all three metrics through a weighted fusion approach:
\begin{equation}
\alpha = w_1 \cdot \alpha_{\text{edge}} + w_2 \cdot \alpha_{\text{variance}} + w_3 \cdot \alpha_{\text{gradient}},
\label{eq:difficulty_fusion}
\end{equation}
where $\alpha \in [0,1]$ is the final difficulty score, and the weights \((w_1,\,w_2,\,w_3)\) are empirically determined via cross-validation using random search optimization~\cite{Bergstra2012}.

\subsubsection{Difficulty-Aware Efficiency Score (DAES)}

To comprehensively evaluate the effectiveness of difficulty-aware routing, we introduce the \textbf{Difficulty-Aware Efficiency Score (DAES)}, which combines accuracy, computational efficiency, and robustness to input complexity:

\begin{equation}
\text{DAES} = \frac{\text{Accuracy} \times \text{Speedup} \times \text{Power\_Efficiency}}{1 + \alpha}
\label{eq:daes}
\end{equation}

Here, $\text{Power\_Efficiency}(m)$ is defined in Eq.~\ref{eq:power_eff}, and $\alpha \in [0,1]$ is the difficulty score computed per input using Eq.~\ref{eq:difficulty_fusion}.

 $\text{Difficulty\_Score} \in [0,1]$ is computed for each test input using our multi-modal difficulty estimation framework (Eq.~\ref{eq:difficulty_fusion}). This ensures a fair complexity-aware evaluation across all methods, including static and non-difficulty-aware baselines, without introducing additional scaling factors.

The DAES metric captures the fundamental trade-off in early-exit networks by jointly considering accuracy, efficiency, and robustness to input complexity. Higher DAES scores indicate models that achieve more balanced performance across heterogeneous input conditions, making the metric particularly relevant for real-world deployment.

\subsection{Joint Exit Policy Optimization}
The core innovation of DART lies in its joint optimization approach that considers all exit points simultaneously, formulated as a global optimization problem to maximize overall efficiency and accuracy trade‐offs.

\subsubsection{Problem Formulation}
Given a neural network with $N$ exits, the optimization objective seeks threshold values $\boldsymbol{\tau} = [\tau_1,\tau_2,\ldots,\tau_{N-1}]$ that maximize:
\begin{equation}
\mathcal{J}(\boldsymbol{\tau})
= \sum_{i=1}^{N} \pi_i(\boldsymbol{\tau}) \bigl[A_i - \beta_{\text{opt}} \cdot C_i\bigr],
\label{eq:objective}
\end{equation}
where $\pi_i(\boldsymbol{\tau})$ is the probability of exiting at layer $i$, $A_i$ is the accuracy achieved when exiting at layer $i$, $C_i$ is the computational cost incurred up to exit $i$, and $\beta_{\text{opt}} \in [0,1]$ is a trade-off parameter that balances accuracy preservation against computational efficiency gains, following dynamic programming principles~\cite{bellman1957dynamic}.

\subsubsection{Dynamic Programming Solution}
We employ a value iteration algorithm over state representations $s = (\text{exit\_index},\alpha_{\text{bin}},\text{confidence}_{\text{bin}})$ to learn optimal exit policies. The Q‐value update combines immediate rewards and future expected returns:
\begin{equation}
Q(s,a) = R(s,a) + \gamma \sum_{s'\in\mathcal{S}} P(s'|s,a)\,V(s'),
\label{eq:q_value}
\end{equation}
where $R(s,a)$ balances accuracy and cost, $\gamma$ is the discount factor, and transitions follow the MDP framework~\cite{watkins1992q}.

\subsubsection{Threshold Calibration}
We generate candidate thresholds using quantiles of the confidence distributions:
\begin{equation}
\tau_i^{\text{candidate}} = \operatorname{quantile}(\mathcal{C}_i, q),
\quad q \in \{0.1,0.2,\ldots,0.9\},
\label{eq:threshold_candidates}
\end{equation}
where $\mathcal{C}_i$ is the confidence distribution at exit $i$, following practices in early‐exit tuning~\cite{zhou2020bert}.

\subsection{Adaptive Coefficient Management}
The adaptive management system continuously refines exit policies through multi‐scale strategies as follows.

\subsubsection{Multi-Strategy Adaptation Framework}
\textbf{Temporal Adaptive Strategy:} Coefficients evolve based on recent performance via exponential decay:
\begin{equation}
c_t = \alpha_{\text{decay}}\cdot c_{t-1} + \bigl(1-\alpha_{\text{decay}}\bigr)\cdot f(\text{performance}_t),
\label{eq:temporal_adaptation}
\end{equation}
where $\alpha_{\text{decay}} \in [0,1]$ is the decay factor, following online learning formulations~\cite{shalev2012online}.

\textbf{Class-Aware Adaptation:} Class‐specific coefficients update based on per‐class performance:
\begin{equation}
c_{\text{class}}^{(t+1)}
= c_{\text{class}}^{(t)}
+ \eta\cdot\bigl(A_{\text{target}} - A_{\text{class}}^{(t)}\bigr),
\label{eq:class_adaptation}
\end{equation}
where $A_{\text{target}}$ is the desired accuracy (e.g., 0.85), $A_{\text{class}}^{(t)}$ is the current accuracy for the specific class (e.g., "car" or "ship" in CIFAR-10), and $\eta$ is the adaptation rate, inspired by meta‐learning updates~\cite{finn2017model}. During deployment, we use the model's high-confidence predictions as pseudo-labels to update class statistics in the absence of ground truth.

\subsubsection{Adaptive Selection and Tracking}
We maintain running statistics over a sliding window of the most recent \(w = 1000\) inferences (per exit and per class): accuracy, confidence distributions, and compute (time/energy). These statistics drive small periodic updates of the adaptive coefficients and thresholds.

When multiple adaptation strategies (e.g., alternative coefficient sets) are available, we select the next strategy using the UCB1 rule to balance exploration and exploitation. UCB1 is a multi-armed bandit algorithm that intelligently chooses between trying new strategies and using proven successful ones, ensuring DART continuously learns optimal coefficient management, as formulated in the following equation:

\begin{equation}
\mathrm{UCB}_i(t)=\bar r_i(t)+\sqrt{\frac{2\ln t}{n_i(t)}},
\label{eq:ucb}
\end{equation}
where \(\bar r_i(t)\) is the windowed reward estimate (accuracy–cost trade-off from Eq.~\ref{eq:objective}) and \(n_i(t)\) counts selections of strategy \(i\). If UCB selection is disabled, the system reduces to deterministic threshold adaptation driven by the same sliding-window statistics.

\subsection{Framework Extensibility to Transformers}

To demonstrate the adaptability of our CNN-focused framework, we extended DART to vision transformers. For LeViT transformers, the framework adapts to token-based representations:

\begin{equation}
\text{ExitBlock}_{\text{ViT}}(T) = \text{MLP}(\text{LayerNorm}(\text{GlobalPool}(T))) \label{eq:vit_exit}
\end{equation}

where $T \in \mathbb{R}^{B \times N \times D}$ represents the token sequence with batch size $B$, $N$ tokens, and $D$ dimensions.

\textbf{Token-Level Difficulty Estimation:} For transformers, we maintain the same difficulty estimation approach as CNNs,
computing α from the input image before tokenization to ensure consistent difficulty assessment across architectures:

\begin{equation}
\alpha_{\text{token}} = w_1 \cdot \alpha_{\text{edge}} + w_2 \cdot \alpha_{\text{variance}} + w_3 \cdot
\alpha_{\text{gradient}} \label{eq:token_difficulty}
\end{equation}

where the difficulty components are computed from the input image using Equations~\ref{eq:edge_density}, \ref{eq:variance},
and \ref{eq:laplacian} before tokenization, with weights $(w_1, w_2, w_3)$.

\subsection{Training and Inference Pipeline}

The training process integrates all components through unified optimization that simultaneously learns network parameters and exit policies.

\subsubsection{Multi-Exit Loss Function}

The total loss combines contributions from all exits with progressive weighting:

\begin{equation}
\mathcal{L}_{\text{total}} = \sum_{i=1}^{N} w_i \cdot \mathcal{L}_{\text{CE}}(y, \hat{y}_i) + \lambda \cdot \mathcal{L}_{\text{policy}} \label{eq:total_loss}
\end{equation}

where $w_i = \frac{i}{N}$ emphasizes later exits. The term $\mathcal{L}_{\text{CE}}$ represents the standard Cross-Entropy loss between the true labels $y$ and the prediction $\hat{y}_i$ from each exit. The term $\mathcal{L}_{\text{policy}}$ is a regularization loss that encourages an efficient exit distribution by penalizing overuse of later exits.

\subsubsection{Inference Optimization}
During inference, the framework applies the learned exit policy while adapting thresholds to the input difficulty. We \emph{increase} the confidence requirement for difficult inputs to avoid premature early exits. Let $\alpha \in [0,1]$ denote the difficulty score of the current input (higher means harder), $\boldsymbol{\tau} \in [0,1]^{N-1}$ the learned base thresholds, and $\boldsymbol{c}\in\mathbb{R}_{+}^{N-1}$ the learned per-exit coefficients. We first apply the coefficients element-wise,
\[
\boldsymbol{\tau}_{\text{adapted}} \;=\; \boldsymbol{c} \odot \boldsymbol{\tau},
\]
and then form the difficulty-aware effective threshold at exit $i$ as
\begin{equation}
\tau_i' \;=\; (\tau_{\text{adapted}})_i \;+\; \beta_{\text{diff}} \cdot \alpha,
\label{eq:inference_threshold}
\end{equation}
followed by clamping to $[0,1]$. Thus, easy inputs ($\alpha \approx 0$) retain near-baseline thresholds, while hard inputs ($\alpha \approx 1$) face higher thresholds and are more likely to continue to deeper exits. We tune the nonnegative sensitivity parameter $\beta_{\text{diff}}\!\ge\!0$ on a validation set to balance responsiveness and stability.

\begin{algorithm}[!t]
\caption{DART Adaptive Exit Decision Algorithm}
\begin{algorithmic}[1]
\STATE \textbf{Input:} Sample $x$, learned thresholds $\boldsymbol{\tau}$, coefficients $\boldsymbol{c}$, difficulty scale $\beta_{\text{diff}}\!\ge\!0$
\STATE Compute difficulty score: $\alpha \leftarrow f_{\text{difficulty}}(x)$
\STATE Apply adaptive coefficients: $\boldsymbol{\tau}_{\text{adapted}} \leftarrow \boldsymbol{c} \odot \boldsymbol{\tau}$
\FOR{$i = 1$ to $N-1$}
\STATE Compute exit prediction: $(\hat{y}_i, \mathrm{conf}_i) \leftarrow E_i(h_i)$
\STATE Difficulty-aware threshold (Eq.~\ref{eq:inference_threshold}):
$\tau_i' \leftarrow (\tau_{\text{adapted}})_i + \beta_{\text{diff}} \cdot \alpha$

\STATE Clamp to $[0,1]$: $\tau_i' \leftarrow \min\!\big(1,\max(0,\tau_i')\big)$
\IF{$\mathrm{conf}_i > \tau_i'$}
\STATE \textbf{return} $\hat{y}_i$, exit\_index $= i$
\ENDIF
\ENDFOR
\STATE \textbf{return} Final prediction from exit $N$
\end{algorithmic}
\end{algorithm}

This pipeline ensures that exit decisions incorporate both learned policies and real-time input characteristics, enabling adaptive inference that responds to input complexity variations.
\section{Experimental Evaluation}

Experiments were conducted to evaluate DART across accuracy, computational efficiency, and adaptive behavior. All experiments were implemented in PyTorch with CUDA acceleration and executed on NVIDIA A100 GPUs. While experiments utilized A100 GPUs, the reported MACs and energy savings are architecture-agnostic metrics that directly translate to efficiency gains on edge accelerators.

Three representative CNN architectures were selected as primary testbeds: AlexNet (8 layers, 61M parameters), ResNet-18 (18 layers, 11M parameters), and VGG-16 (16 layers, 138M parameters). MNIST and CIFAR-10 were used as two standard datasets for evaluation. To further examine framework generality, experiments were extended to LeViT-128s, a transformer-based model with 7.8M parameters, serving as a proof-of-concept for transformer applications.

The difficulty estimation module used empirically determined weights of $(w_1 = 0.4, w_2 = 0.3, w_3 = 0.3)$ for fusing edge density, pixel variance, and gradient complexity metrics. The difficulty-aware threshold adaptation used $\beta_{\text{diff}} = 0.3$ as the sensitivity parameter. The adaptive coefficient manager employed an exponential decay strategy with $\alpha_{\text{decay}} = 0.95$.

Performance was benchmarked against two baselines, including static models without early exits, and BranchyNet with fixed thresholds \cite{teerapittayanon2016branchynet}. Evaluation metrics included top-1 accuracy, computational cost in multiply–accumulate (MAC) operations, wall-clock inference time, and energy consumption measured via NVIDIA-SMI. In addition, exit distributions were analyzed to assess routing patterns and adaptive behavior.

\subsubsection{Metric Definitions}

All latency, energy, and power measurements were obtained under identical conditions (same batch size, precision, dataloader, and warmup). Each method was executed $R$ times, and the median latency was used to compute derived metrics.

Speedup relative to static baselines was calculated as:
\begin{equation}
\text{Speedup}(m) = \frac{T_{\text{Static}}}{T_m},
\end{equation}
where $T_m$ is the per-inference wall-clock time of method $m$.
Average power was defined as:
\begin{equation}
P_m = \frac{E_m}{T_m},
\end{equation}
with energy $E_m$ obtained by integrating instantaneous power during inference.
Finally, normalized power efficiency was reported as:
\begin{equation}
\text{Power\_Efficiency}(m) = \frac{E_{\text{Static}}}{E_m},
\label{eq:power_eff}
\end{equation}
with $\text{Power\_Efficiency}(\text{Static})=1.0\times$ by definition.

\subsection{Performance Comparison}

Table~\ref{tab:merged_performance} summarizes the results across all models and datasets.

\begin{table*}[!t]
\caption{Performance and Difficulty-Aware Analysis (MNIST $\alpha\!=\!0.76$; CIFAR-10 $\alpha\!=\!0.85$).}
\label{tab:merged_performance}
\centering
\scriptsize
\setlength{\tabcolsep}{2pt}
\renewcommand{\arraystretch}{1.1}
\begin{tabular}{@{}ll c ccc c c c@{}}
\toprule
\textbf{Architecture} & \textbf{Method} & \textbf{Acc.(\%)} & \textbf{Time(ms)} & \textbf{Energy(mJ)} & \textbf{Power(W)} & \textbf{Speedup} & \textbf{Power Eff.} & \textbf{DAES} \\
\midrule
\multicolumn{9}{c}{\textit{MNIST Results}} \\
\midrule
AlexNet & Static & 98.97 & 0.09 & 5.90 & 65.56 & $1.0\times$ & 1.00 & 0.562 \\
        & BranchyNet & 99.13 & 0.08 & 5.27 & 65.88 & $1.13\times$ & 1.12 & 0.713 \\
        & RL-Agent & 99.49 & 0.06 & 2.29 & 38.20 & $1.50\times$ & 2.57 & 2.179 \\
        & DART & 99.31 & 0.03 & 1.15 & 38.33 & $3.0\times$ & 5.13 & 8.684 \\
\midrule
\multicolumn{9}{c}{\textit{CIFAR-10 Results (CNNs)}} \\
\midrule
AlexNet & Static & 85.29 & 0.08 & 5.18 & 64.75 & $1.0\times$ & 1.00 & 0.461 \\
        & BranchyNet & 83.15 & 0.07 & 4.57 & 65.29 & $1.14\times$ & 1.13 & 0.579 \\
        & RL-Agent & 84.42 & 0.05 & 1.84 & 36.73 & $1.60\times$ & 2.82 & 1.888 \\
        & DART & 82.86 & 0.05 & 1.88 & 37.60 & $1.60\times$ & 2.76 & 1.978 \\
\midrule
ResNet-18 & Static & 88.32 & 0.27 & 17.66 & 65.41 & $1.0\times$ & 1.00 & 0.477 \\
          & BranchyNet & 87.72 & 0.16 & 10.43 & 65.19 & $1.69\times$ & 1.69 & 1.354 \\
          & RL-Agent & 87.87 & 0.14 & 8.10 & 57.84 & $1.93\times$ & 2.18 & 1.998 \\
          & DART & 85.35 & 0.12 & 7.53 & 62.75 & $2.25\times$ & 2.35 & 2.439 \\
\midrule
VGG-16 & Static & 79.16 & 0.20 & 9.41 & 47.05 & $1.0\times$ & 1.00 & 0.428 \\
       & BranchyNet & 81.82 & 0.11 & 3.70 & 46.25 & $1.82\times$ & 2.54 & 4.63 \\
       & RL-Agent & 80.89 & 0.10 & 3.91 & 39.13 & $2.00\times$ & 2.40 & 2.021 \\
       & DART & 80.20 & 0.06 & 2.54 & 42.33 & $3.33\times$ & 3.71 & 5.356 \\
\midrule
\multicolumn{9}{c}{\textit{CIFAR-10 Results (LeViT Transformers)}} \\
\midrule
LeViT-128S & Static & 95.80 & 11.55 & 750.75 & 65.00 & $1.0\times$ & 1.00 & 0.518 \\
           & DART & 81.73 & 4.56 & 150.10 & 32.92 & $2.53\times$ & 5.00 & 5.588 \\
\midrule
LeViT-192 & Static & 96.91 & 19.94 & 1296.10 & 65.00 & $1.0\times$ & 1.00 & 0.524 \\
          & DART & 80.33 & 5.57 & 259.18 & 46.53 & $3.58\times$ & 5.00 & 7.772 \\
\midrule
LeViT-256 & Static & 97.28 & 62.55 & 4065.75 & 65.00 & $1.0\times$ & 1.00 & 0.526 \\
          & DART & 86.11 & 18.89 & 813.30 & 43.05 & $3.31\times$ & 5.00 & 7.704 \\
\bottomrule
\end{tabular}
\end{table*}

DART consistently improves efficiency across CNNs. On CIFAR-10, it achieves a 1.60× speedup on AlexNet (82.86\% vs 85.29\% accuracy, 2.7× energy reduction), a 2.25× speedup on ResNet-18 (85.35\% vs 88.32\% accuracy, 2.3× energy reduction), and a 3.33× speedup on VGG-16 (80.20\% vs 79.16\% accuracy, 3.7× energy reduction). On MNIST, AlexNet with DART reaches 3.0× speedup and 5.1× energy reduction while slightly improving accuracy (99.31\% vs 98.97\%).

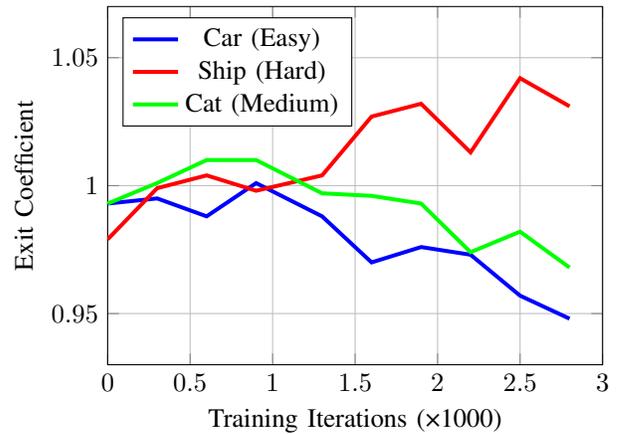
\begin{figure}[!t]
\centering
\begin{tikzpicture}
\begin{axis}[
    xlabel={Training Iterations (×1000)},
    ylabel={Exit Coefficient},
    width=0.45\textwidth,
    height=0.35\textwidth,
    legend pos=north west, 
    grid=major,
    xmin=0, xmax=3.0,
    ymin=0.93, ymax=1.07
]
\addplot[color=blue, line width=1.5pt] coordinates {
    (0.0, 0.993) (0.3, 0.995) (0.6, 0.988) (0.9, 1.001) (1.3, 0.988) (1.6, 0.970) (1.9, 0.976) (2.2, 0.973) (2.5, 0.957) (2.8, 0.948)
};
\addlegendentry{Car (Easy)}
\addplot[color=red, line width=1.5pt] coordinates {
    (0.0, 0.979) (0.3, 0.999) (0.6, 1.004) (0.9, 0.998) (1.3, 1.004) (1.6, 1.027) (1.9, 1.032) (2.2, 1.013) (2.5, 1.042) (2.8, 1.031)
};
\addlegendentry{Ship (Hard)}
\addplot[color=green, line width=1.5pt] coordinates {
    (0.0, 0.993) (0.3, 1.001) (0.6, 1.010) (0.9, 1.010) (1.3, 0.997) (1.6, 0.996) (1.9, 0.993) (2.2, 0.974) (2.5, 0.982) (2.8, 0.968)
};
\addlegendentry{Cat (Medium)}
\end{axis}
\end{tikzpicture}
\caption{Evolution of adaptive coefficients during training for three CIFAR-10 classes based on real experimental data.}
\label{fig:adaptive_behavior}
\end{figure}
For LeViT transformers, DART achieves 2.53–3.58× speedups but with accuracy reductions of up to 17 percentage points, indicating that CNN-oriented early-exit strategies are not directly transferable to attention-based models.

\subsection{Difficulty-Aware Performance}
Observing CIFAR-10 inputs, it is evident that they exhibit moderate complexity with a mean difficulty score of $\alpha \approx 0.85$, consistent across models, confirming that difficulty reflects dataset characteristics rather than architectural bias.

Table~\ref{tab:merged_performance} reports DAES values alongside standard performance metrics. Three insights emerge.
First, CNNs benefit substantially: for example, VGG-16 improves from 0.428 (static) to 5.356 (DART), a 12.5× gain.
Second, transformers also exhibit improvements: LeViT-128S improves from 0.518 to 5.588 (10.8×), LeViT-192 from 0.524 to 7.772 (14.8x), and LeViT-256 from 0.526 to 7.704 (14.6×), showing that DART’s principles generalize beyond CNNs.
Third, across all architectures, DAES improvements confirm that DART exploits input complexity effectively to achieve favorable efficiency–accuracy trade-offs.

\begin{table}[!t] \caption{Extensibility Study: LeViT Transformer Performance Analysis on CIFAR-10} \label{tab:levit_detailed} \centering \scriptsize \setlength{\tabcolsep}{4pt} \renewcommand{\arraystretch}{1.1} \resizebox{\columnwidth}{!}{%
\begin{tabular}{@{}lcccccc@{}} \toprule \textbf{Model} & \textbf{Method} & \textbf{Acc. (\%)} & \textbf{MACs (M)} & \textbf{Time (ms)} & \textbf{Speedup ($\times$)} \\ \midrule LeViT-128S & Static & 95.80 & 282.1 & 11.55 & \(1.00\times\) \\ & DART & 81.73 & 56.4 & 4.56 & \(2.53\times\) \\ \midrule LeViT-192 & Static & 96.91 & 601.1 & 19.94 & \(1.00\times\) \\ & DART & 80.33 & 120.2 & 5.57 & \(3.58\times\) \\ \midrule LeViT-256 & Static & 97.28 & 1053.3 & 62.55 & \(1.00\times\) \\ & DART & 86.11 & 210.7 & 18.89 & \(3.31\times\) \\ \bottomrule \end{tabular}%
}
\end{table}

To validate the efficiency of DART’s difficulty-estimation module against the state of the art, the computational overhead is compared against RACENet~\cite{racenet} using FLOPs and latency. Only the control mechanisms responsible for dynamic behavior are considered to ensure a fair overhead comparison, i.e., DART’s input-difficulty estimator and RACENet’s class-aware adaptive normalization. Latency values were measured on an NVIDIA GPU with a batch size of 128, averaged over 5,000 runs to capture per-sample throughput. DART’s \texttt{DifficultyEstimator} adds 78.9K FLOPs for lightweight input analysis, whereas RACENet’s adaptive normalization requires a dedicated MLP at every layer, contributing 716,912 additional parameters and 3.96M FLOPs. This results in RACENet incurring \textbf{50.3$\times$} higher compute overhead and a substantially larger memory footprint than DART, clearly highlighting its inefficiency for deployment under resource constraints

\subsection{Adaptive Behavior}

Figure~\ref{fig:adaptive_behavior} illustrates the evolution of exit coefficients during training. The adaptive system learns class-specific strategies: for easy classes (car), coefficients decrease from 0.99 to 0.95, enabling more aggressive early exits. For hard classes (ship), coefficients increase from 0.98 to 1.05, resulting in more conservative exits. Medium classes (cat) show intermediate evolution (0.99 to 0.97). This demonstrates the framework’s ability to adapt policies dynamically based on real data distributions.

\subsection{CNN Architecture Analysis}

A detailed analysis of CNNs highlights architecture-dependent benefits. AlexNet achieves the highest power savings (42\%), while ResNet-18 and VGG-16 show distinct accuracy efficiency trade-offs, underscoring the importance of architecture-aware tuning. The results confirm that DART’s joint optimization converges to stable exit distributions, effectively balancing accuracy and efficiency across multiple exits while learning meaningful exit strategies.

\subsection{ViT Architecture Analysis}

The extensibility study with LeViT models provides insight into applying DART to non-CNN architectures. While efficiency gains are consistent (2.5–3.6× speedups, substantial energy reduction), accuracy drops remain noticeable (up to 17 points). This limitation highlights fundamental differences between CNN and transformer representations: early transformer layers primarily capture token positioning and structural cues rather than semantic features, making early exits more detrimental than in CNNs. These results point to the need for transformer-specific early-exit mechanisms, such as attention-aware difficulty metrics, token-level exit strategies, and specialized calibration.

Overall, the experimental results confirm that DART achieves consistent improvements across CNNs, with speedups of up to 3.33× and energy savings exceeding 5× while maintaining competitive accuracy. The framework adapts effectively to varying input complexities, learns class-dependent strategies in real time. The findings demonstrate the practicality of DART as a unified framework for dynamic neural network optimization on resource-constrained edge AI platforms.

\vspace{-0.29cm}
\section{Conclusion}

This paper introduces \textbf{DART} (Input-Difficulty-Aware Adaptive Threshold), a framework that enhances early-exit neural networks by tackling suboptimal exit policies, missing input-difficulty awareness, and independent threshold optimization. DART integrates three innovations: (1) a lightweight difficulty estimation module with minimal overhead, (2) a joint exit policy optimization via dynamic programming, and (3) an adaptive coefficient management system. Experiments on AlexNet, ResNet-18, and VGG-16 show up to \textbf{3.3$\times$} speedup, \textbf{5.1$\times$} lower energy, and \textbf{42\%} lower average power versus static networks, with competitive accuracy. Extending DART to Vision Transformers (LeViT) yields power (5.0$\times$) and execution-time (3.6$\times$) gains but also accuracy loss (up to 17 percent), underscoring the need for transformer-specific early-exit mechanisms. Finally, the proposed Difficulty-Aware Efficiency Score (DAES) demonstrates up to \textbf{14.8$\times$} improvement over baselines, capturing DART’s superior accuracy–efficiency–robustness trade-offs.

\section{Acknowledgements}
\scriptsize 
This work was supported in part by the Estonian Research Council grant PUT PRG1467 "CRASHLESS“, EU Grant Project 101160182 “TAICHIP“, by the Deutsche Forschungsgemeinschaft (DFG, German Research Foundation) – Project-ID "458578717", and by the Federal Ministry of Research, Technology and Space of Germany (BMFTR) for supporting Edge-Cloud AI for DIstributed Sensing and COmputing (AI-DISCO) project (Project-ID "16ME1127").

\bibliographystyle{IEEEtran}
\bibliography{ref}

\end{document}